\documentclass[twocolumn,showpacs]{revtex4}
\usepackage{color}
\usepackage{graphicx}
\usepackage{dcolumn}
\usepackage{amsmath}

\begin{document}
\bibliographystyle{revtex}

\title[Short Title]{Coulomb forces in neutron star crusts}

\author{C. O. Dorso and P. A. Gim\'enez Molinelli}

\affiliation{Departamento de F\'isica, FCEN, Universidad de Buenos
Aires, N\'u\~nez, Argentina}
\affiliation{IFIBA-CONICET}

\author{J. A. L\'opez and E. Ram\'irez-Homs}

\affiliation{Department of Physics, University of Texas at El
Paso, El Paso, Texas 79968, U.S.A.}

\date{\today}

\begin{abstract}
We study the role the proton-electron gas interaction has on the formation of nuclear structures in neutron star crusts.   Using a classical molecular dynamics model we study isospin symmetric and asymmetric matter at subsaturation densities and low temperatures varying the Coulomb interaction strength.  The effect of such variation is quantified on the fragment size multiplicity, the inter-particle distance, the isospin content of the clusters, the nucleon mobility and cluster persistence, and on the nuclear structure shapes.  We find that the proton-electron Coulomb interaction distributes matter more evenly, disrupts the formation of larger objects, reduces the isospin content, modifies the nucleon average displacement, but does not affect the inter-nucleon distance in clusters .  The nuclear structures are also found to change shapes by different degrees depending on their isospin content, temperature and density.
\end{abstract}

\maketitle

\section{Introduction}\label{intro}

The {\it raison de vivre} of neutron stars is a balance between nuclear, Coulomb and gravitational forces.  In vacuum, the weak decay $n\rightarrow p+e^-+\bar{\nu}_e$ is exothermic with a net energy release of $0.778 \ MeV$ as kinetic energies of the emitted particles.  In conditions of thermal and chemical equilibrium, the opposite process, $p+e\rightarrow n+\nu$, which is endothermic, would work to restore the original ratio of protons and neutrons, this, however, is not the case at neutron star pressures and densities.  Due to Pauli principle arguments, the neutron decay is inhibited and such processes yield to an accumulation of neutrons; the neutron dominance is expected to start in the stellar crust when $Z/A\sim 0.13$~\cite{horo-2006}, while the outer layers maintain larger neutron to proton ratios in interesting structures dictated by the varying percentage of protons and neutrons in an all-embedding electron gas.

The electron gas is expected to play an important role in the formation of the nuclear structures.  As the density increases as a function of depth in the crust, the electrons will transform protons into neutrons by inverse $\beta$ decay making the neutron-to-proton ratio more asymmetric and forming neutron-rich nuclei.  In addition to being responsible for producing neutrons, the electron cloud also creates a Coulomb potential that surrounds the rest of the matter contributing to define its shape.

At densities and temperatures expected to exist in neutron star crusts ($ \rho \lesssim \rho_0$ and $T\lesssim 1.0 \ MeV$, with $ \rho_0$ denoting the normal nuclear density), nucleons form structures that are substantially different from the quasi-spherical nuclei we are familiar with in our part of the universe.  Such structures, which have been dubbed ``nuclear pasta'', have been investigated using various models~\cite{20, 21, 22, 23, 24, 25, 26, 27, 28, 29, 30, Dor12} which have shown them to be the result of the interplay of nuclear and Coulomb forces in an infinite medium.  The structure of the nuclear pasta is expected to play an important role in the study of neutrino opacity in neutron stars \cite{horo_lambda}, neutron star quakes and pulsar glitches~\cite{32}.

In a previous study, a combination of classical molecular dynamics and fragment recognition algorithms~\cite{Dor12} was shown to be very effective in the study of the pasta structures.  By using a microscopic view of the nucleons, it was found that a combination of topological tools can be used to classify the different structures into recognizable patterns; this opens the door to cross-model comparison between structures obtained with different approaches and to a quantifiable analysis of the effect the nuclear and Coulomb forces have on the pasta formation.  This investigation will focus on the role of the Coulomb forces on the creation of the pasta.

The presence of the electron gas is bound to have, at least, a two-fold effect.  On one side, its interaction with the protons should certainly modify their positioning in space as well as their kinetic energy and, on another hand, the fact that the gas fills all openings should reduce the long range proton-proton interactions with an screening-like effect.

In a $2003$ investigation~\cite{30}, the screening effect of an electron gas on cold nuclear structures was investigated using a static liquid-drop model, and it was found that main effect of the gas screening was to extend the range of densities where bubbles and clusters appear.  Indeed, the screening was found to be of minor importance although the study, being static, did not include any spatial and dynamical effects the gas should have produced on the nucleons.

A $2005$ study~\cite{Maruyama-2005} used a density functional method to investigate charge screening on nuclear structures at subnuclear densities but still at zero temperature; in particular, cases with and without screening were directly compared. The main results of the study were to extract the nucleon density profiles and to quantify the spatial rearrangement of the proton and electron charge densities.  Once again, it was found that the density region in which the pasta exists becomes broader when the Coulomb screening is taken into account, mainly due to the rearrangement of the protons; the authors remark the importance of extending such study to finite temperatures and with dynamical models.

Recently, in an extension of previous works~\cite{horo_lambda,P14,P15}, Piekarewicz and Toledo S\'anchez~\cite{P2012} performed Monte Carlo simulations directly on $5,000$ nucleons to determine the lowest proton fractions that are compatible with $\beta$ equilibrium in neutron star crust environments; a feature that depends finely on the Coulomb interaction between the protons and electron gas.  The study, which approximated the Coulomb interaction via an Ewald summation, was performed at a single temperature ($T=1.0 \ MeV$) and it succeeded in finding a relationship between the proton fractions and the density of the electronic contribution.  In closing, the authors recommended that further investigations be performed with molecular dynamics simulations to study dynamical observables and to examine the role of the temperature in the study of stellar crusts.

Thus, the motivation of this work is to study the effect the proton-electron gas interaction has on the formation of the nuclear pasta using a dynamical model, at non-zero temperatures, and in a systematic way that takes advantage of the tools developed in our previous work~\cite{Dor12}.  In the next section the model used (classical molecular dynamics) is briefly introduced along with the cluster recognition algorithms and analysis tools employed.  Section~\ref{Coulomb} presents a series of results of the simulations and analyses performed.  We summarize some of our findings in Section~\ref{concluding} and present a quick look at work in progress.

\begin{figure}[h]  
\begin{center}
\includegraphics[width=2.5in]{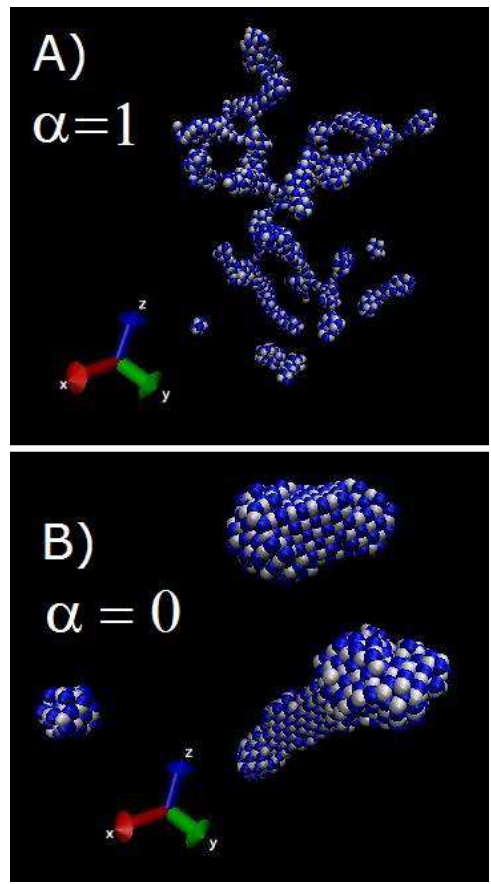}
\end{center}
\caption{(Color online) Characteristic structures obtained with ($\alpha$=$1$) and without ($\alpha$=$0$) the proton-electron gas interaction for symmetric matter ($x=0.5$) at density $\rho=0.015 \ fm^{-3}$ and temperature $T=0.1 \ MeV$.} \label{x5t1d015}
\end{figure}

\section{Model used}\label{cmd}
The model used here was developed to study nuclear reactions from a microscopic point of view, and it is composed of a molecular dynamics code retrofitted with cluster recognition algorithms and a plethora of analysis tools.  The justification for using this model in stellar crust environments was presented elsewhere~\cite{Dor12}, here we simply mention some basic ingredients of the model.

The classical molecular dynamics model ($CMD$), first introduced in~\cite{14a}, has been used in heavy-ion reaction studies to explain experimental data~\cite{Che02}, identify phase transitions and other critical phenomena~\cite{16a,Bar07,CritExp-1,CritExp-2}, explore the caloric curve of nuclear matter~\cite{TCalCur,EntropyCalCur} and isoscaling~\cite{8a,Dor11}.  Synoptically, $CMD$ uses two two-body potentials to describe the motion of ``nucleons'' by solving their classical equations of motion.  The potentials were developed phenomenologically by Pandharipande~\cite{pandha}:
\begin{eqnarray*}
V_{np}(r) &=&V_{r}\left[ exp(-\mu _{r}r)/{r}-exp(-\mu
_{r}r_{c})/{r_{c}}
\right] \\
& &\ \mbox{}-V_{a}\left[ exp(-\mu _{a}r)/{r}-exp(-\mu
_{a}r_{c})/{r_{c}}
\right] \\
V_{NN}(r)&=&V_{0}\left[ exp(-\mu _{0}r)/{r}-exp(-\mu _{0}r_{c})/{
r_{c}}\right] \ , \label{2BP}
\end{eqnarray*}
where $V_{np}$ is the potential between a neutron and a proton, and $V_{NN}$ is the repulsive interaction between either $nn$ or $pp$. The cutoff radius is $r_c=5.4$ $fm$ and for $r>r_c$ both potentials are set to zero. The Yukawa parameters $\mu_r$, $\mu_a$ and $\mu_0$ were determined by Pandharipande to yield an equilibrium density of $\rho_0=0.16 \ fm^{-3}$, a binding energy $E(\rho_0)=16$ MeV/nucleon and a selectable compressibility of  $250 \ MeV$ for the ``Medium'' model, and $535 \ MeV$ for the ``Stiff'''~\cite{pandha}. We remark that, since the potential $V_{NN}$ does not permit bound states between identical nucleons, pure neutron matter is unbound and, at a difference from potentials used by other models~\cite{horo_lambda}, the Pandharipande potentials have a hard core.

The main advantage of the $CMD$ model is the possibility of knowing the position and momentum of all particles at all times as this allows the study of the structure of the nuclear medium from a microscopic point of view.  The outcome of $CMD$, namely, the time evolution of the particles in $(\mathbf{r},\mathbf{p})$, can be used as input in anyone of the several cluster recognition algorithms that some of us have designed for the study of nuclear reactions~\cite{Dor95,Str97,dor-ran}.

As explained elsewhere~\cite{Lop00,Dor12,piekatesis}, the lack of quantum effects, such as Pauli blocking, --perhaps the only serious caveat in classical models-- ceases to be relevant in conditions of high density and temperature (such as in heavy-ion reactions) or in the low-density and low-temperature stellar environments.

To simulate an infinite medium, systems with thousands of nucleons were constructed using $CMD$ under periodic boundary conditions. Cases symmetric in isospin (i.e. with $x=Z/A=0.5$, $1000$ protons and $1000$ neutrons), and cases with asymmetric isospin (with $x=0.3$ comprised of $1000$ protons and $2000$ neutrons) were constructed in cubical boxes with sizes adjusted to have densities between $\rho=0.01 \ fm^{-3} \le \rho \le  \rho_0$; our previous study, \cite{Dor12}, presented an {\it sm\"org{\aa}sboard} of structures obtained through this method.

\begin{figure}[b]  
\begin{center}
\includegraphics[width=3.55in]{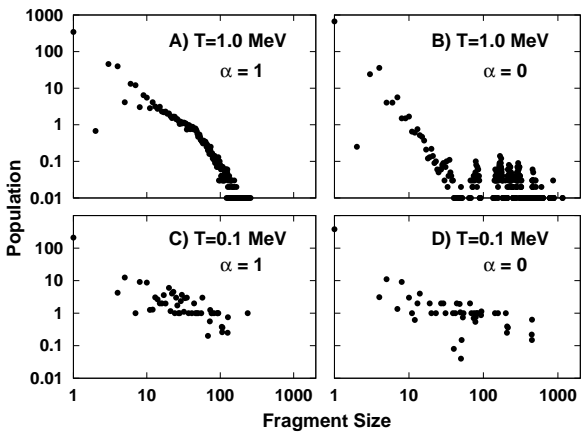}
\end{center}
\caption{Effect of the screened potential on the fragment size multiplicity.  Plots show the distribution of cluster sizes observed in 200 configurations of $x=0.3$ nuclear matter at density $\rho=0.015 \ fm^{-3}$ and temperatures $T=1.0 \ MeV$ (top) and $T=0.1 \ MeV$ (bottom). The figure on the left panel correspond to configurations with Coulomb interaction, and those on the right to the case without such potential.} \label{massDistx3D015}
\end{figure}

\subsection{Coulomb interaction}\label{cp}

The neutron star crust is embedded in a degenerate electron gas.  Although the nucleon-electron system is overall neutral and $\beta$-equilibrated, to take into account the Coulomb interaction (which is of infinite range) it is necessary to use some approximation.  The two most common approaches are the Thomas-Fermi screened Coulomb potential (used, e.g., in $QMD$~\cite{26}) and the Ewald summation procedure~\cite{wata-2003}; $CMD$ has been used with both approximation (see~\cite{Dor12} for a comparison of methods under $CMD$) but for computational reasons we opt for the use of a screened Coulomb potential.

\begin{figure}[t]  
\begin{center}
\includegraphics[width=3.2in]{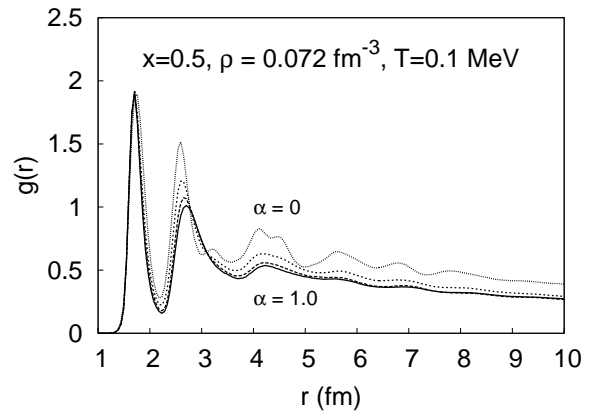}
\end{center}
\caption{Examples of the radial correlation function for varying strengths of the Coulomb potential: $\alpha=1$ (full Coulomb), $0.8$, $0.2$, and $0$ (without Coulomb).}\label{Radial}
\end{figure}

We approximate the electron gas as a uniform ideal Fermi gas with the same number density as the protons.  Its inclusion in the $CMD$  is through the use of a screened Coulomb potential obtained from the Poisson equation, $V_C^{(Scr)}(r)=({e^2}/{r})\exp(-r/\lambda)$, where the screening length,  $\lambda$, is related to the electron mass, Fermi momentum, and number density.  Here, pragmatically, and after testing various lengths, we use the prescription used before~\cite{26,horo_lambda} and set $\lambda=10 \ fm$, which satisfies the requirement of being sufficiently smaller than the size of the simulation cell, $L=\left( A/\rho\right)^{ 1/3 }$.

\begin{figure}[t]  
\begin{center}$
\begin{array}{cc}
\includegraphics[width=3.4in]{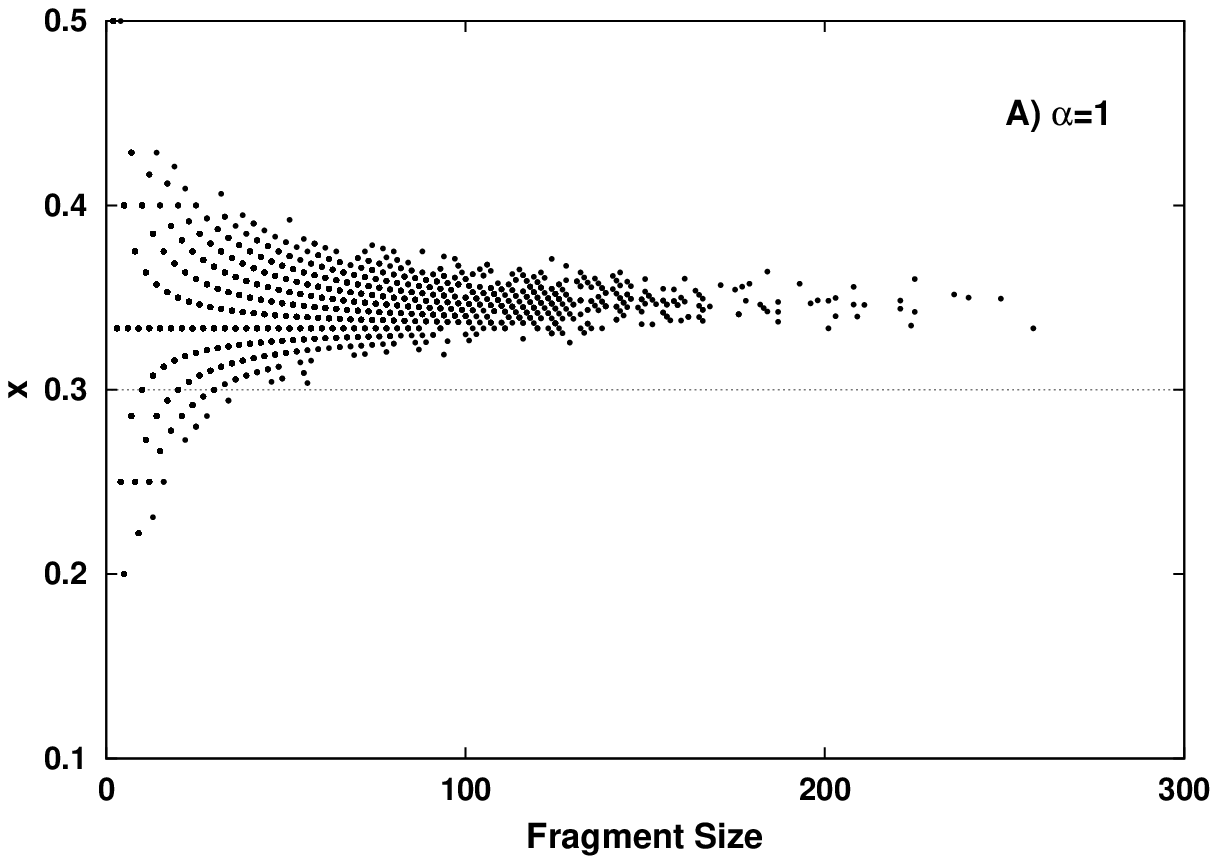} \\
\includegraphics[width=3.4in]{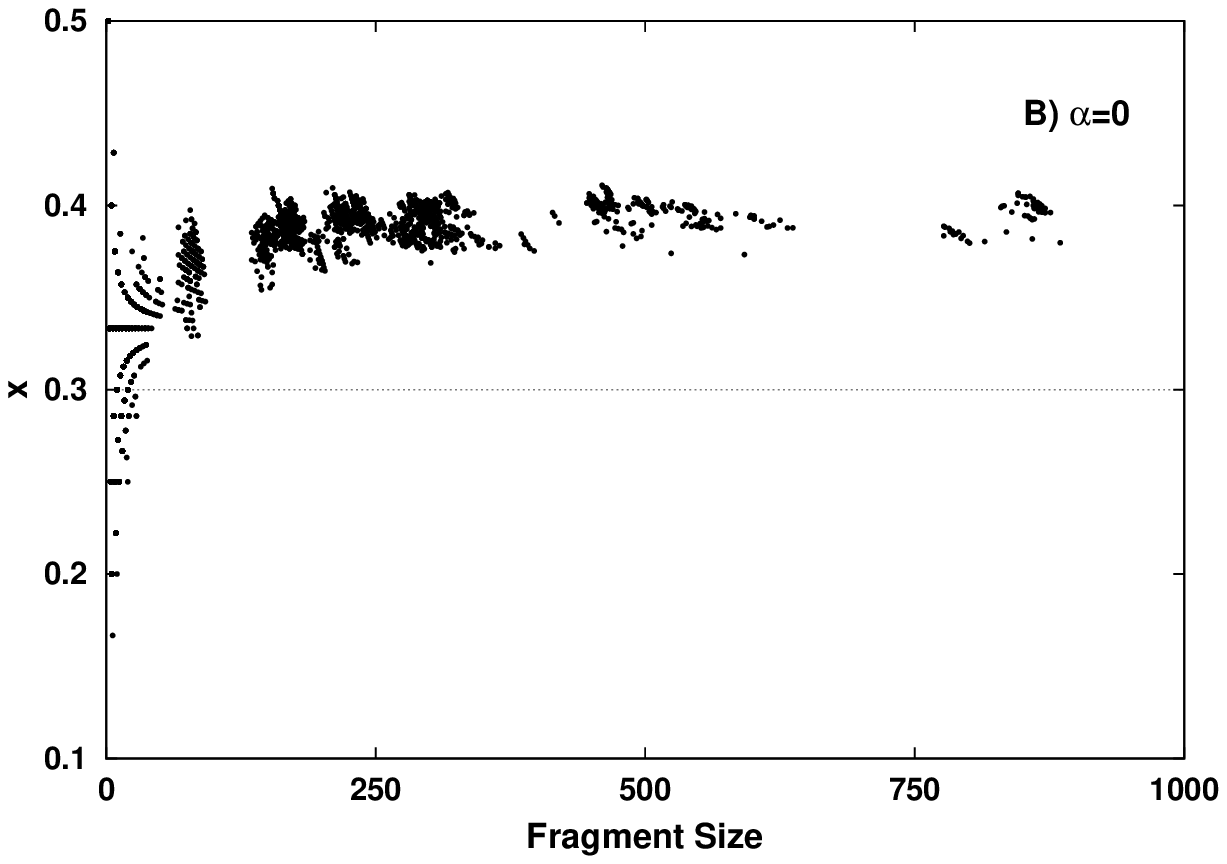}
\end{array}$
\end{center}
\caption{$x$ content of the clusters formed in $200$ configurations of asymmetric matter ($x=0.3$) at density $\rho=0.015 \ fm^{-3}$ and temperature $T=1.0 \ MeV$. The top panel shows the case with Coulomb and the bottom one the case without such interaction. Notice that the abscissas have different scales.} \label{XfragX3D015T1}
\end{figure}

\subsection{Simulation procedure}\label{procedure}

The trajectories of the nucleons are then governed by the Pandharipande and the screened Coulomb potentials.  The equations of motion are solved using a symplectic Verlet integration with energy conservation of $\mathcal{O}$($0.01 \%)$. The nuclear system is force-heated or cooled using isothermal molecular dynamics with the Andersen thermostat procedure~\cite{andersen} with small temperature steps.  To simulate stellar crust conditions we operate in the range of $0.1\le T \le 1.0 \ MeV$ and densities $\rho=0.01 \ fm^{-3} \le \rho \le  \rho_0$; in comparison to heavy-ion reactions, nucleons in neutron star crusts are practically in a semi-frozen state.

To obtain reliable statistics we sample each $T$, $\rho$ and $x$ configuration $200$ times in an stationary ergodic process. That is, after reaching equilibrium at a given set of values of $T$, $\rho$ and $x$, the $(\mathbf{r},\mathbf{p})$ values of all nucleons are stored for further processing while the system continues its evolution for a ``long'' time ($4000$ time steps) until it reaches a state independent of the previous one, point at which the procedure is repeated for a total of $200$ times.  The recorded nucleon positions and momenta are then used later to identify clusters and to characterize the structure as explained next.
\begin{figure}[h]  
\begin{center}
\includegraphics[width=3.in]{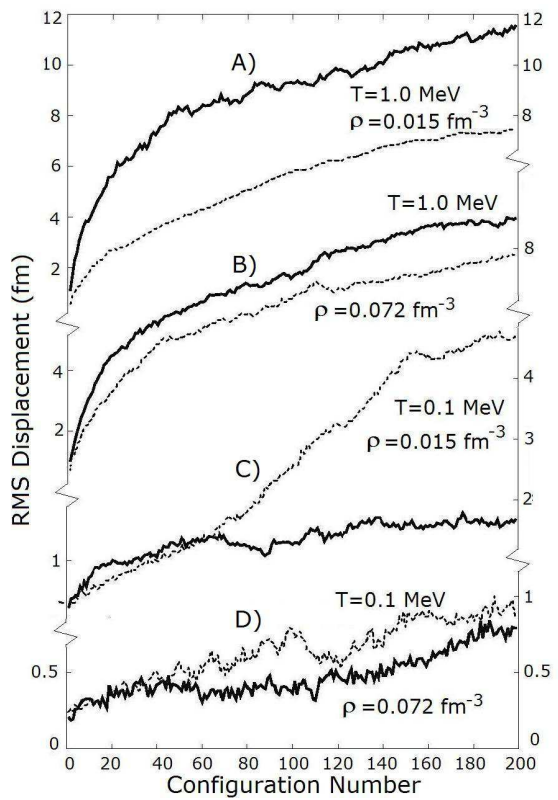}
\end{center}
\caption{$RMS$ displacement of nucleons during the evolution of the stationary ergodic process with Coulomb (continuous line) and without Coulomb (dashed line) as labeled; see text for details.  Notice that at low temperatures the mobility with Coulomb is smaller than without Coulomb, while at higher temperatures the effect is the opposite.} \label{desplx5T1}
\end{figure}

Given that, for the densities and temperatures of interest, the nucleons do not have large mobilities, the task of identifying self-bound cumuli is sufficiently simple as to be achieved with a basic ``Minimum Spanning Tree'' ($MST$) algorithm~\cite{Dor95,Str97}.  The $MST$ finds cluster membership simply by identifying those particles that are closer to each other than some clusterization radius, which here is set to $3.0 \ fm$ based on the range of the nuclear potential and densities investigated; the generic method was modified to recognize fragments that extend into any of the adjacent periodic cells.

\subsection{Analysis tools}\label{analysis}

Once the fragments have been identified, their global properties are studied as follows.  $MST$ yields information about the fragment multiplicity; in Ref.~\cite{Dor12} systems at $x=0.3$ and $0.5$ and at various densities ($0.01 \ fm^{-3}\le \rho \le 0.015 \ fm^{-3}$) and temperatures ($0.1 \ MeV \le T\le 1.0 \ MeV$) exhibited increases in multiplicity with temperature (in opposition to the rise-fall that takes place in intermediate-energy heavy-ion reactions) as well as the formation of percolating structures that encompass most of the nucleons in the simulation.

At a microscopic level, the dynamics of the nucleons in such systems can be quantified through their average displacement as a function of ``time'', i.e. through the time steps of the simulation; for the cases mentioned before (cf.~\cite{Dor12}) the mobility between configurations was found to be between $2$ and $20 \ fm$.  Likewise, the microscopic stability of the clusters can be gauged through the ``persistency''~\cite{DorStr95,Lop00} which measures the tendency of members of a given cluster to remain in the same cluster; for the cases previously studied, colder clusters ($T\lesssim0.5 \  MeV$) exchange less than half of their nucleons during 200 ergodic configurations, while those at higher temperatures ($T  \gtrsim 0.8 \ MeV$) replace up to $90\%$ of their constituents.

Another interesting descriptor is the isospin content of each cluster produced, i.e. their $x$ values; in the previous study it was found that for the case of $x=0.3$, small clusters ($A\lesssim10$) tend to have less protons than the global average, while such effect was not present in the case of $x=0.5$.  Other global characterization tool is the pair correlation function, $g(r)$, which gives information about the spatial ordering of the nuclear medium by comparing the average local density to the global density; in the previous study $g(r)$ showed that nucleons in clusters have an interparticle distance of about $1.7 \ fm$ at all studied densities, and was useful in determining the onset of {\it lasagna}-like structures.

\begin{figure}[h]  
\begin{center}
\includegraphics[width=3.in]{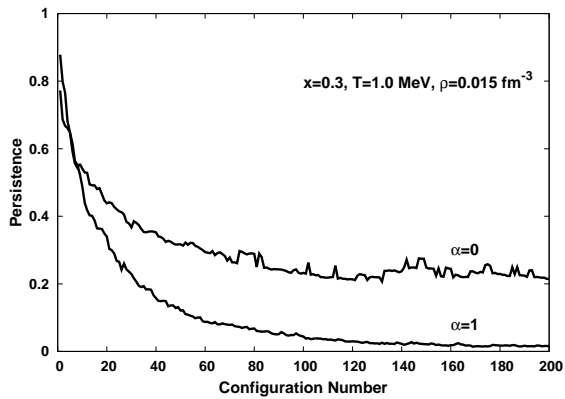}
\end{center}
\caption{Persistence with and without Coulomb as it evolves through 200 configurations of the stationary ergodic process.} \label{persis}
\end{figure}

Proceeding beyond global measures, the shapes of nuclear structures can be characterized by their volume, surface area, mean curvature, plus an interesting construct known as the Euler characteristic; these four objects comprise what is known as the ``Minkowsky functionals'' which completely describe all topological properties of any three-dimensional object.  The evaluation of the mean curvature (averaged over the whole surface of the object) and Euler number (a topological invariant connected to the vertices, edges and faces of polyhedra), however, requires the mapping of the nuclear clusters into a polyhedra; procedure that can be accomplished through the Michielsen--De Raed algorithm~\cite{michielsen,29}.  

In~\cite{Dor12} it was shown that generic structures, such as ``gnocchi'', ``spaghetti'', ``lasagna'' and ``crossed-lasagnas'' or ``jungle gym'' and their inverse structures (with voids replacing particles and viceversa), all have well defined and distinct values of the mean curvature and Euler numbers with magnitudes dictated by the overall size of the structure, i.e. by the number of particles used.  It should be remarked that the assignment of curvature and Euler values to a given structure becomes relative to the overall scale.  This becomes important for structures with near-zero Euler number which signal spaghetti, lasagna and their anti-structures; curled spaghetti rods of a given length, for instance, can easily oscillate between positive and negative near zero Euler numbers by curling or uncurling a segment.  This classification, which was respected at all $x$ values, densities and temperatures studied, is shown in Table~\ref{table1}.

It is with this model and arsenal of tools that we now embark on a study of the effect the proton-electron gas interaction has on the formation of the nuclear pasta.

\begin{table}[ht]
\centering  
\caption{Classification Curvature - Euler}
\begin{tabular}{c|| c | c | c} 
\hline                  
& Curvature $<0$ & Curvature $\sim 0$ & Curvature $>0$ \\ 
\hline \hline                        
Euler $>0$ & Anti-Gnocchi &  & Gnocchi  \\
Euler $\sim0$ & Anti-Spaghetti & Lasagna & Spaghetti  \\
Euler $<0$ & Anti-Jungle Gym &  & Jungle Gym  \\ [1ex]      
\hline 
\end{tabular}
\label{table1} 
\end{table}

\section{The effect of Coulomb} \label{Coulomb}
Along the line of previous studies~\cite{30,Maruyama-2005,P2012} and adhering to their advise, in this study we use a molecular dynamics model to study spatial and dynamical effects the electron gas may produce on the nuclear structures at non-zero temperatures.  Specifically we extend the approach of one of the previous studies and compare pasta structures obtained with and without the proton-electron gas interaction as well as with varying strengths of it.

In summary, pasta structures were obtained at subsaturation densities  ($0.015 \ fm^{-3} \le \rho \le 0.072 \ fm^{-3}$) and low temperatures ($0.1 \ MeV \le T \le 1.0 \ MeV$) for symmetric ($x=0.5$) and asymmetric ($x=0.3$) cases; altogether $200$ simulations were carried out per each combination of $\{\rho,T,x\}$.  To see the effect of the Coulomb interaction, each case was ``cooked'' repeatedly with a screened Coulomb potential with a varying amplitude, i.e. with $\alpha V_C^{(Scr)}$, where $0 \le \alpha \le 1$.  For a fair comparison, all corresponding cases with different values of $\alpha$ were produced with identical initial conditions of $\{x,\rho,T\}$ and started off from the same initial random configuration.  As an illustration, Figure~\ref{x5t1d015} shows two corresponding structures obtained with and without Coulomb.

The effect of the screened potential can also be observed on the fragment size multiplicity.  Figure~\ref{massDistx3D015} shows the distribution of cluster sizes observed in 200 configurations of asymmetric matter ($x=0.3$) at density $\rho=0.015 \ fm^{-3}$ and temperatures $T=0.1 \ MeV$ and $T=1.0 \ MeV$. The figures corresponds to configurations with (cases A and C) and without (B and  D) the Coulomb interaction.  The increase of the number of fragments of sizes $A \gtrsim 500$ as $\alpha$ goes from $\alpha=1$ to $\alpha=0$ underline the role the Coulomb force has on determining the mass distribution.

The change of the inner structure can be quantified through the use of the radial distribution function, $g(r)$.  An example of this effect is shown in Figure~\ref{Radial} which shows this function calculated with the structures obtained with symmetric nuclear matter ($x=0.5$) at density $\rho=0.072 \ fm^{-3}$ and temperature $T=0.1 \ MeV$ and for varying strengths of the Coulomb potential, namely $\alpha=1$ (full Coulomb), $0.8$, $0.2$, and $0$ (without Coulomb).  As it can be easily observed, the proton-electron gas interaction does not affect the nearest neighbor distances, although the magnitude of the function for second nearest neighbors increases as Coulomb weakens.

The isotopic content on the cluster is also affected by the presence of the proton-electron gas interaction.  Figure~\ref{XfragX3D015T1} shows the $x$ content of the clusters formed in $200$ configurations of asymmetric matter ($x=0.3$) at density $\rho=0.015 \ fm^{-3}$ and temperature $T=1.0 \ MeV$. Again, the top panel corresponds to the case with full Coulomb and the bottom one to the case without such interaction.  Clearly visible is the enhancement up to $x\approx 0.4$ that would occur without Coulomb; also noticeable is the increase of the maximum fragments sizes which grow up to $A\approx 800$ without the electric interaction.

As suspected in~\cite{Maruyama-2005}, in this study we find that the interactions between protons and electrons are responsible for the rearrangement of protons, although we find this to be of varying degrees.  Figure~\ref{desplx5T1} shows the $RMS$ displacement of nucleons as a function of the evolution of the stationary ergodic process.  Each curve is made of 200 points, each of which represents the average displacement (with respect to the original configuration) of the $2000$ nucleons used in the $x=0.5$ cases shown.  For each case tracked ($\rho=0.015$ and $0.72 \ fm^{-3}$ at $T=0.1$ and $1.0 \ MeV$), the displacements obtained with and without Coulomb are presented.  Curiously, at low temperatures Coulomb seems to enhance the nucleon mobility much more than at high temperatures.

A similar effect is observed in the persistence, which is the tendency of nucleons in a given fragment to remain in the same cluster.  Figure~\ref{persis} shows the persistence with and without Coulomb as it evolves through 200 configurations of the stationary ergodic process, measured with respect to the original configuration of $3000$ nucleons with  $x=0.3$, $\rho=0.015$ and $T=1.0 \ MeV$.  As seen in the mobility, Coulomb enhances the transfer of nucleons thus decreasing the persistence as shown by the two curves.

\begin{figure}[t]  
\begin{center}$
\begin{array}{cc}
\includegraphics[width=3.4in]{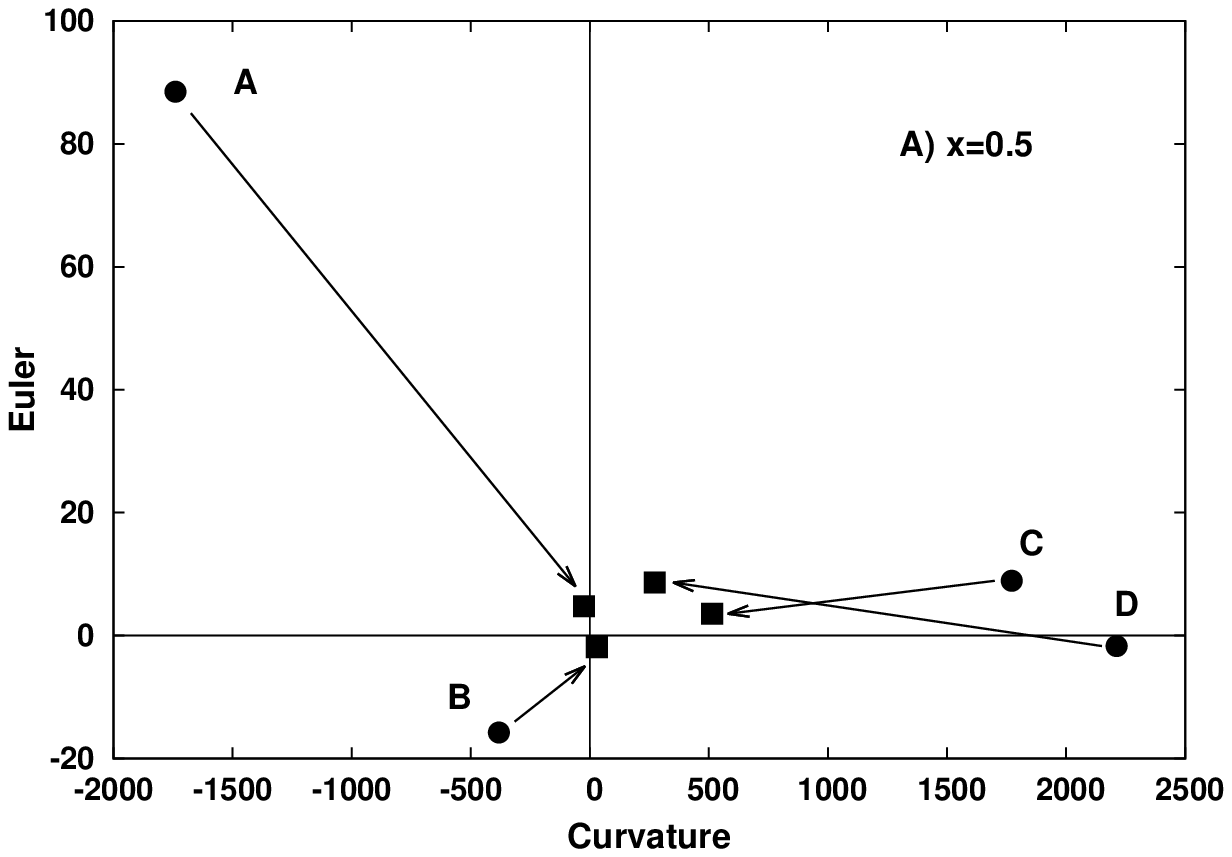} \\
\includegraphics[width=3.4in]{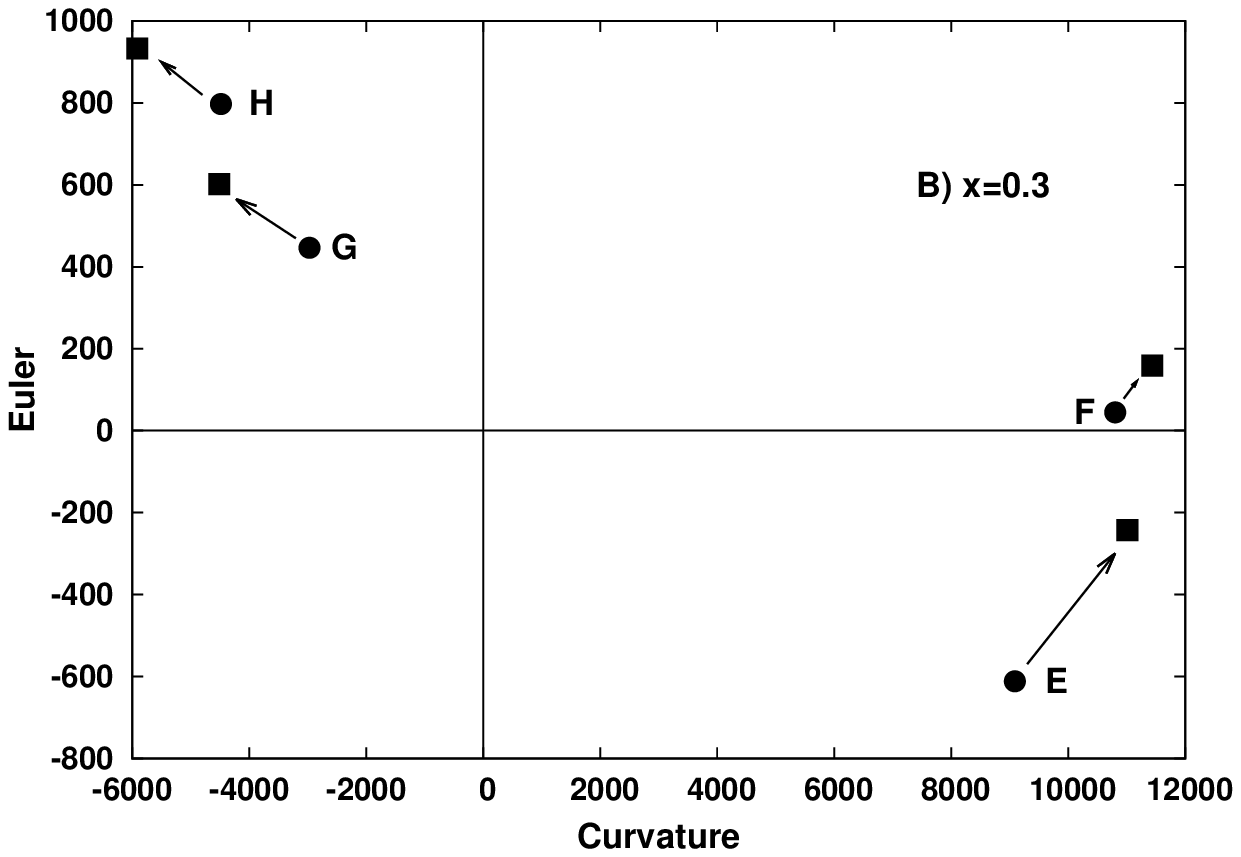}
\end{array}$
\end{center}
\caption{Average values of the Curvature and Euler numbers of the structures listed in
Table~\ref{table2}; circles correspond to structures with Coulomb and squares to structures without Coulomb, arrows indicate the average displacement of the structures as $\alpha$ goes from $1$ to $0$. } \label{curv-euler}
\end{figure}

The effect of Coulomb on the shape of the structures formed can be quantified using the Euler Characteristic and Mean Curvature introduced before.  Following the procedure explained in more detail elsewhere~\cite{Dor12}, digitized polyhedra were constructed for each of the 200 nuclear structures obtained with and without Coulomb at each value of $\{x,\rho,T\}$ considered.  The corresponding values of the mean curvature and Euler numbers were calculated  and, although the configurations corresponding to the same set of conditions were not identical, their curvature and Euler numbers showed robust average values.  Table ~\ref{table2} shows these values and Figure~\ref{curv-euler} their location on the Curvature-Euler plane as well as their changes in position as the Coulomb strength is diminished.  The standard deviations of the points are, roughly speaking, of the size of the points used in the plots.

\begin{table}[ht]
\centering  
\caption{Classification Curvature - Euler}
\begin{tabular}{c|c|c|cc|cc} 
\hline \hline                        
 & $\rho$ & $T$ & \multicolumn{2}{c|}{$\alpha=1$} & \multicolumn{2}{c}{$\alpha=0$} \\
& ($fm^{-3}$) & ($MeV$)  & Curvature & Euler & Curvature & Euler \\
[0.5ex]
A & 0.072 & 1.0 & -1740	& 89 & -24 	& 4.7 \\
B & 0.072 & 0.1 & -382 &	-15.8 &	 29.4	&  -1.9 \\
C & 0.015 & 0.1 & 1771 &	8.9 &	 514 &	 3.5 \\
D & 0.015 & 1.0 & 2212 &	-1.7	& 273 &	 8.6 \\
E & 0.015 & 1.0 & 9090 & -612 & 11011 & -243 \\
F & 0.015 & 0.1 & 10804 & 454	& 11436 &	 159 \\
G & 0.072 & 1.0 & -2969 & 447 & -4512 & 602 \\
H & 0.072 & 0.1 & -4485 & 798 & -5914 & 933 \\
[1ex]      
\hline 
\end{tabular}
\label{table2}
\end{table}

\subsection{Analysis}

The results obtained in this study allow us to understand the effect the proton-electron gas interaction has on the formation of the nuclear pasta. Perhaps the most interesting result of this investigation is the fact that the characteristic topological structures of the  nuclear pasta --which are usually associated with the presence of the competing interactions of the long range Coulomb potential and the short range nuclear force-- exist even when there is no Coulomb (i.e. when $\alpha \rightarrow 0$).  We believe this to be the result of the competition between attractive and repulsive nuclear forces at very low temperatures; in our case this is due to the attraction of the $V_{np}$ and the repulsion of the $V_{nn}$ and $V_{pp}$ which operate at medium ranges; incidentally, we find the same effect also in other potentials~\cite{P2012} as will be reported in a future communication.

Thus the role of the Coulomb force is to emphasize such competition between attractive-repulsive forces.  The examples of structures formed with and without Coulomb (Figure~\ref{x5t1d015}) suggest that the electric forces tend to distribute matter more and form less-compact objects.  This behavior is in agreement with the reduction of the fragment size multiplicity undoubtedly induced by Coulomb as illustrated in Figures~\ref{massDistx3D015} and~\ref{XfragX3D015T1}.

At a microscopic scale, Figure~\ref{Radial} shows that varying the strength of the Coulomb interaction does not change the inter-particle distance, but certainly decreases a bit the $x$ content of the fragments as can be seen in Figure~\ref{XfragX3D015T1}.  In agreement with this, the increment of nucleon mobility produced by Coulomb (cf. Figure~\ref{desplx5T1}) gets reflected in a reduction of the persistence (cf. Figure~\ref{persis}).

With respect to the modification of the shapes due to the electric interaction, we find that Coulomb tends to reduce both the curvature and Euler number of the structures.  Indeed, Coulomb appears to ``compactify'' symmetric matter into spaghetti and lasagna type structures as shown clearly in Figure~\ref{curv-euler} for symmetric matter (top panel); of course, the effect becomes blurrier in systems with smaller charge densities, such as the asymmetric matter case of $x=0.3$ (bottom panel).

\section{Concluding remarks}\label{concluding}

The effect of the proton-electron gas interaction in symmetric and neutron-rich matter was studied at densities and temperatures of interest to neutron star crusts by varying the interaction strength.  Its effect on the fragment size multiplicity, the inter-particle distance, the isospin content of the clusters, the nucleon mobility, and on the modification of the topological shape due to the Coulomb interaction was studied.

The most general result is the existence of the  nuclear pasta structures even without the presence of the Coulomb interaction due to the interplay between the attractive and repulsive parts of the nuclear forces.  In general, the proton-electron Coulomb interaction tends to distribute matter more, form less-compact objects, increase nucleon mobility, reduce the persistence and the fragment size multiplicity, decrease the $x$ content of the clusters, but does not change the inter-particle distances.  All of this gets reflected on a modification of the shapes of the nuclear structures formed by a reduction of their curvature and Euler numbers; effect that is more pronounced in symmetric than asymmetric matter.

In conclusion, we find that the electron gas screening is of upmost importance in the formation of the nuclear pasta in neutron star crust environments.  In a future work we will extend the present study to investigate the transition from a uniform nuclear phase to the pasta structures which is bound to occur at very low temperatures, as well as its dependence on the isotopic content and compressibility of the nuclear matter using both versions of the Padharipande potential as well as other classical potentials~\cite{P2012, dor88}.

\begin{acknowledgments}
C.O.D. is supported by CONICET Grant PIP0871, P.G.M. by a CONICET scholarship, and J.A.L. and E.R.H. by NSF-PHY Grant 1066031.  C.O.D., J.A.L. and E.R.H. thank the hospitality of the Nuclear Theory Group of the Nuclear Science Division of The Lawrence Berkeley National Laboratory where this work was performed. The three-dimensional figures were prepared using the software {\it Visual Molecular Dynamics}~\cite{VMD}.
\end{acknowledgments}

\end{document}